# Suppression of the ferromagnetic state in LaCoO$_3$ films by rhombohedral distortion

D. Fuchs[1], L. Dieterle[2], E. Arac[1,3], R. Eder[1], P. Adelmann[1], V. Eyert[4], T. Kopp[4], R. Schneider[1],

D. Gerthsen[2], and H. v. Löhneysen[1,3]

[1]*Forschungszentrum Karlsruhe, Institut für Festkörperphysik, 76021 Karlsruhe, Germany*

[2]*Laboratorium für Elektronenmikroskopie, 76128 Karlsruhe, Germany*

[3]*Physikalisches Institut, Universität Karlsruhe, 76128 Karlsruhe, Germany*

[4]*Center for Electronic Correlations and Magnetism, Universität Augsburg, 86135 Augsburg, Germany*

Epitaxially strained LaCoO$_3$ (LCO) thin films were grown with different film thickness, $t$, on (001) oriented (LaAlO$_3$)$_{0.3}$(SrAl$_{0.5}$Ta$_{0.5}$O$_3$)$_{0.7}$ (LSAT) substrates. After initial pseudomorphic growth the films start to relieve their strain partly by the formation of periodic nano-twins with twin planes predominantly along the <100> direction. Nano-twinning occurs already at the initial stage of growth, albeit in a more moderate way. Pseudomorphic grains, on the other hand, still grow up to a thickness of at least several tenths of nanometers. The twinning is attributed to the symmetry lowering of the epitaxially strained pseudo-tetragonal structure towards the relaxed rhombohedral structure of bulk LCO. However, the unit-cell volume of the pseudo-tetragonal structure is found to be nearly constant over a very large range of $t$. Only films with $t > 130$ nm show a significant relaxation of the lattice parameters towards values comparable to those of bulk LCO.

Measurements of the magnetic moment indicate that the effective paramagnetic moment, $m_{eff}$, and thus the spin state of the $Co^{3+}$ ion does not change for films with $t \leq 100$ nm. However, the saturated ferromagnetic moment, $m_s$, was found to be proportional only to the pseudo-tetragonal part of the film and decreases with increasing rhombohedral distortion. The measurements demonstrate, that ferromagnetism of LCO is strongly affected by the rhombohedral distortion while the increased unit-cell volume mainly controls the effective paramagnetic moment and thus the spin state of the $Co^{3+}$ ion.

# I. Introduction

The perovskite-type lanthanum cobaltate LaCoO$_3$ (LCO) has recently attracted much attention due to its unusual electronic and magnetic properties at ambient pressure[1,2] and the observation of ferromagnetism in epitaxially strained thin films.[3,4] At low temperature, $T \leq 35$ K, LCO is a nonmagnetic semiconductor with a ground state of Co$^{3+}$ ions in a low-spin (LS) configuration ($t_{2g}^6 e_g^0$, $S = 0$).[5,6] This is believed to change to a primarily intermediate-spin (IS) ($t_{2g}^5 e_g^1$, $S = 1$) state[7] in the temperature range 35 K $< T <$ 100 K, and further to a mixture of IS and high-spin (HS) ($t_{2g}^4 e_g^2$, $S = 2$) states in the interval 300 K $< T <$ 600 K. The crossover between spin states with increasing temperature arises from a delicate interplay between the crystal-field splitting, $\Delta_{CF}$, between the $t_{2g}$ and $e_g$ energy levels, and the intra-atomic exchange interaction (Hund`s rule coupling), $\Delta_{EX}$. The balance between $\Delta_{CF}$ and $\Delta_{EX}$ can be affected by, e. g., hole or electron doping[8] and by chemical or external pressure.[9,10] Since $\Delta_{CF}$ is very sensitive to changes of the Co-O bond length, $d$,[11] and the Co-O-Co bond angle, $\gamma$, structural changes with respect to both easily modify the spin state of the Co$^{3+}$ ion. We have shown that the population of higher spin states is enhanced in tensile strained LCO films.[3] Calculations based on the generalized gradient approximation to the density functional theory indeed show that a magnetic state is more stable than the LS state for structures with larger Co-O distances and more open Co-O-Co bond angles.[12]

In contrast to the nonmagnetic LS ground state of bulk LCO ferromagnetic (FM) order below $T_c \approx 85$ K is found for epitaxially strained thin films.[3] Zhou *et al.* also reported on FM order of LCO nanoparticles below 85 K.[13] The experimental results support the assumption that a reduced Jahn-Teller distortion may play a crucial role in the stabilization of the FM state. However, $T_c$ values around 85 K have also been observed by Yan *et al.* and more recently by Harada *et al.* reporting on ferromagnetism at the surface of bulk LCO material.[14,15] It was proposed that ferromagnetism of LCO is likely related to hole doping with oxygen by chemisorption on the surface,[16] which leads to a tetrahedral coordination of the HS $Co^{2+}$ ion by $O^{2-}$ ions on the surface. In a very recent work, Herklotz *et al.*[17] have observed a reversible strain effect on the ferromagnetic state of LCO films grown epitaxially on piezoelectric substrates. These results strongly indicate that the ferromagnetic state in the strained LCO films is not caused by chemisorption but rather by a structural distortion. Generalized-gradient-corrected density functional calculations by Ravindran *et al.*[18] showed that the rhombohedral distortion stabilizes the nonmagnetic ground state in LCO. In addition, spin-polarized calculations on the hypothetical cubic pervoskite phase of LCO showed that the FM phase is lower in energy than the corresponding nonmagnetic phase. The origin of the magnetism has been explained in terms of itinerant ferromagnetism. The lowering of the total energy is due to an energy gain by an exchange splitting of the $t_{2g}$ levels. The analysis of the electronic structure showed that a Peierls-Jahn-Teller-like instability arises in the FM cubic phase which leads to the rhombohedral distortion in LCO. The rhombohedral distortion enhances (reduces) the hybridization between the $t_{2g}$ ($e_g$) states and the O $2p$ states which leads to a band broadening (narrowing) and therefore to a reduction of the $t_{2g}$ derived density of states at the Fermi energy, $N(E_F)$. This results in an almost linear decrease of the magnetic moment. Therefore, they explained the increasing trend of the magnetic moment and the stabilization of the IS state of $Co^{3+}$ as a function of Sr doping in $La_{1-x}Sr_xCoO_3$ by the narrowing of the bands, i. e., the increase of $N(E_F)$, due to the increase of the volume.

The calculations are consistent with our previously published results on strained LCO films. The effective magnetic moment, $m_{eff}$, in the paramagnetic state increases with the cube root of the unit-cell volume, $V^{1/3}$. The increase of $V$ is caused by the elastic properties of LCO which has been grown on different substrate materials with different lattice matching. The Curie temperature, $T_c$, increases also strongly with increasing tensile strain and seems to saturate at $T_c = 85$ K. The increase of $T_c$ and $m_{eff}$ has been related to a continuous increase of the Co-O-Co bond angle, $\gamma$, i. e., a decrease of the octahedral site rotation, which leads to a reduced hybridization of Co $3d$ derived $t_{2g}$ and O $2p$ states.

In this work we present new experimental results on the growth and relaxation process of epitaxially strained LCO films, which evidence the suppression of the FM state in LCO films with increasing rhombohedral distortion. The experiments are consistent with density functional calculations and contribute to a better understanding on the FM state in LCO.

**II. Experimental**

LCO films were prepared with varying film thickness ranging from 8 nm up to 220 nm on (001) oriented LSAT substrates. A detailed description of the film preparation has already been given in Refs. 3,4. Here, we only refer in short to the film preparation carried out by pulsed laser deposition at substrate temperatures of about 650°C using stoichiometric sinter targets of LCO. After deposition the samples were cooled down to 500°C where they were kept for 30 min in 0.9 bar oxygen atmosphere to ensure a complete and homogeneous oxygenation of the films.

For the determination of the structural properties of the epitaxial films such as the out-of-plane *c*-axis lattice parameter and the mosaic spread we carried out $\theta/2\theta$ scans and rocking curves on a two-circle x-ray diffractometer in Bragg-Brentano geometry. A four-circle x-ray diffractometer was used to determine the in-plane lattice parameters. The film thickness was obtained from measurements with a Dektak profilometer, cross-section transmission electron microscopy (TEM) images, and x-ray diffraction where thickness oscillations were visible.[19] The chemical composition was checked by Rutherford-backscattering spectrometry (RBS) and energy dispersive x-ray absorption analysis (EDX) and found to be highly stoichiometric with respect to the cations.

The microstructure of plan-view and cross-section samples was studied by TEM. Selected-area electron diffraction (SAED), conventional and high-resolution TEM (CTEM/HRTEM) imaging was performed with a 200 keV Philips CM200 FEG/ST equipped with a field-emission gun. TEM cross-section samples were prepared by gluing a sandwich, sawing, polishing, dimpling with rotating polishing felt (with 3 µm, 1 µm and 0.25 µm diamond paste) and Ar-ion milling as outlined in Ref. [20]. Surface amorphization was removed efficiently by low-energy Ar-ion etching at 200 eV. High-resolution TEM lattice-fringe images were evaluated by strain-state analysis using the digital analysis of lattice images (DALI) software[21] where lattice-parameter or displacements are measured with respect to the known lattice parameter of the substrate (in our case LSAT).

The magnetic properties of the films were studied using a superconducting quantum interference device (SQUID) system from Quantum Design. In order to determine $T_c$ and $m_{eff}$, field-cooled (FC) magnetization measurements were carried out in the temperature range $4\,\text{K} \leq T \leq 150\,\text{K}$. The external field strength of $\mu_0 H = 20$ mT was applied parallel to the film surface. The diamagnetic contribution of the substrate was determined in separate runs and subtracted from the magnetization data. To deduce the saturated moment in the FM state, $m_s$, magnetization measurements as a function of external field were done up to $\mu_0 H = 7$ T at $T = 10$ K.

**III. Results and Discussion**

A. Structural relaxation of epitaxial LCO films – structure, rhombohedral distortion, and twinning.

Bulk LCO crystallizes in a rhombohedral perovskite structure with the space group $R\bar{3}c$ and the lattice parameters $a_{rh} = 5.3788$ Å and $\alpha_{rh} = 60.798°$. The oxygen atoms are twisted around the [111] direction. The displacements of the oxygen atoms with respect to an idealized cubic structure can be specified by the distortion parameter $\delta x = 0.049$.[22] Within the unit cell the Wyckoff positions of the atoms are thus as follows: La 2$a$ (¼, ¼, ¼), Co 2$b$ (0, 0, 0) and O 6$e$ (¼-$\delta$x, ¼+$\delta$x, ¾). It is convenient to describe the rhombohedral perovskite structure of LCO in terms of a pseudo-cubic unit-cell whose axes correspond to those of the aristotype (ideal perovskite), with the lattice parameter $a_{pc}$ and angle $\alpha_{pc}$ (pseudocubic description is marked with the index $pc$ in the following), see Fig. 1 a). The rhombohedral phase is formed by a compression along one of the four body diagonals of the idealized cubic unit cell, <111> resulting in four different rhombohedral variants. A combination by any two of them form one twin, sketched in Fig. 1 b), which results in four different domain states, forming (100)$_{pc}$ or (110)$_{pc}$ twin-planes. With the general dot product for non-rectangular coordinate systems the lattice parameter $a_{pc}$ and angle $\alpha_{pc}$ can be calculated to $a_{pc} = \frac{1}{2}\sqrt{3 - 2\cos\alpha_{rh}} \cdot a_{rh}$ = 3.826 Å and

$$\alpha_{pc} = \arccos(\frac{1 - 2\cos\alpha_{rh}}{2\cos\alpha_{rh} - 3}) = 90.69°.$$

However, the experimental value for the O-Co-O bond angle, $\eta$, amounts to a somewhat larger value of $\eta$ = 91.2° at room temperature[23,24]. The deviation is likely caused by the irregularity of the CoO$_6$ octahedra, i. e., different Co-O bonding lengths, caused by the Jahn-Teller distortion[25] and thus, related to the rhombohedral distortion.

The rhombohedral distortion leads to a decrease of the Goldschmidt tolerance factor $G = d_{A-O}/\sqrt{2} d_{B-O} <$ 1 in the $AB$O$_3$-type perovskite, where $d_{A-O}$ and $d_{B-O}$ are the La-O and Co-O bond lengths, respectively. The distortion corresponds to a cooperative rotation of the $B$O$_6$ octahedra by an angle $\phi$ around the [111]$_{pc}$ direction, reducing the symmetry from cubic to rhombohedral. A noticeable distortion of the octahedra does not occur if the rotations around any of the threefold axes of the regular octahedra are small.[26] For regular, undistorted octahedra the rotation angle $\phi$ can be derived from the rhombohedral angle $\alpha_{rh}$ by the relation cos $\alpha_{rh}$ = (4-cos$^2\phi$)/(4+2cos$^2\phi$) resulting in a rotation of $\phi \approx$ 10.3°.[23,27]

With increasing tolerance factor towards the perfect cubic structure ($G = 1$), $\phi$, $\alpha_{rh}$ and $\alpha_{pc}$ decrease towards 0°, 60° and 90°, respectively. This is achieved by increasing $d_{A-O}$ by, e. g., doping with Sr (La$_{1-x}$Sr$_x$CoO$_3$) for x > 0.5 or by increasing the temperature due to the thermal expansion. A reduction of $\phi$ may likewise be expected by an expansive lattice strain. Pervoskites such as LCO and related materials like LaAlO$_3$ undergo a phase transition from the rhombohedral perovskite structure $R\overline{3}c$ to a cubic perovskite structure at higher temperatures, e. g., $T = 1340$°C for LCO.[28-30]

Most phase transitions to lower symmetry are accompanied by twin formation. Symmetry breaking at the phase transition gives rise to energetically equivalent domain states in the low-temperature phase which are mapped onto each other by the symmetry elements which have been lost with respect to the prototype symmetry.[31] In particular, pervoskites such as LaAlO$_3$ or LCO which undergo a displacive phase transition from a paraelastic cubic perovskite at high temperatures to a ferroelastic rhombohedral structure usually show ferroelastic domains formed by deformation twinning.[30,32,33]

The LCO films were grown on (001) oriented LSAT single-crystal substrates with a lattice constant of $a_s$ = 3.865 Å (Ref. [34]). Hence, the epitaxial LCO films experience an in-plane tensile lattice strain, due to the lattice mismatch of $\varepsilon = (a_{pc} - a_s)/a_s$ = -1%. All films that we have grown showed a single-phase (001)$_{pc}$ oriented cube-on-cube growth, i. e., the in-plane directions of the film [100]$_{pc}$ and [010]$_{pc}$ are nearly parallel to [100] and [010] of the substrate, respectively. In the initial stage of growth the films usually show a pseudomorphic growth mode, where the film adopts structural properties and symmetry of the substrate material within the growth plane with an in-plane film lattice parameter $a_f = a_s$.

The high-resolution TEM (HRTEM) cross section micrograph of an 8 nm thick LCO film on (001) oriented LSAT, Fig. 2 a), visualizes the abrupt and flat interface between epitaxial film and substrate.

The image shows uniform contrast without the formation of dislocations, antiphase boundaries, domains, or intermediate layers. Perfect coherence of the lattice planes crossing the interface suggests that the misfit accommodation is purely elastic. Beside such an elastic accommodation of substrate-induced stress, perovskites also often show interfacial dislocations, the formation of antiphase domains or even phase transitions to relieve epitaxial strain[35]. $(002)_{pc}$ lattice-fringe images (not shown here) obtained by the interference of the transmitted beam and the strongly excited $(002)_{pc}$ reflection show an intensity modulation perpendicular to the growth direction with a periodicity of 12±2 nm. The result of a strain-state analysis shown in Fig. 2 c) indicates that the growth is not pseudomorphic in a strict sense. In Fig. 2 c), the displacements of the $(001)_{pc}$ planes in the LCO film are measured with respect to a reference lattice with the lattice parameter of the LSAT substrate. They are displayed in a color-coded scale in units of the LSAT (001)-plane distance. Slight variations in the LSAT substrate (yellow and red) are attributed to surface effects induced by the TEM sample preparation. Compressive strain is observed generally in the LCO film along the [001] direction. However, a sinusoidal modulation is revealed in addition producing small rhombohedrally distorted regions which merge fluently into another without the formation of twins as schematically shown in Fig. 2 d). This modulation can be considered as a precursor stage of the nano-twinning observed in LCO films thicker than 15 nm.

The bright-field micrograph of a plan-view sample with an 8 nm LCO film imaged in $(220)_{pc}$ two-beam condition, Fig. 2 b), exhibits domains with a striped contrast and stripe orientations along the [100] and [010] directions and an average stripe distance of about 15 nm. This slightly larger modulation periodicity compared to the periodicity derived from cross-section TEM images can be explained by stress relieve of LCO due to the small thickness of the LSAT substrate after the plan-view sample preparation. The lateral size of the domains is between 50 nm and 200 nm. High strain and brittleness of thicker LCO films impedes the preparation of plan-view samples because fracture occurs during the preparation process.

Fig. 3 a) shows a $(002)_{pc}$ lattice-fringe image of a 50 nm LCO film in a cross-section perspective. Bright and dark diffuse contrast bands perpendicular to the interface with an average periodicity of 19±4 nm are marked by white arrows. Twin formation is visualized in Fig. 3 b) where the region displayed in Fig. 3 a) is compressed by a factor of five along the $[100]_{pc}$ direction. Black lines in Fig. 3 b) emphasize the inclination of the $(002)_{pc}$ planes of the film with respect to the LSAT substrate which is expected after twin formation. The rhombohedral distortion of LCO film with $t \geq 50$ nm is confirmed by selected area electron diffraction which shows the expected rhombohedral splitting of the $(4\ 0\ 4)_{pc}$ and $(\bar{4}\ 0\ 4)_{pc}$ LCO reflections. LCO films with t ≥ 50 nm contain dislocations and cracks in contrast to thinner films.

Because of the negative lattice mismatch and the pseudo-cubic structure of LSAT the films adopt a pseudo-tetragonal structure with $a_f \approx a_s > c_f$, where $c_f$ is the out-of-plan film lattice parameter.[4] Due to the nearly perfect and defect-free pseudomorphic growth, the mosaic spread of such films compares to that of the single-crystal substrates. A non-elastic structural relaxation usually comprises structural lattice distortions which lead to an increase of the mosaic spread, $\Delta\omega$, of the film. $\Delta\omega$ was measured by the full width at half maximum (FWHM) of the rocking curves at $(00l)_{pc}$ film reflections. Fig. 4 shows the rocking curves at the $(002)_{pc}$ reflection of LCO films deposited on (001) LSAT for different film thicknesses, ranging from 8 – 100 nm. Besides the central peak around $\omega_0 \approx 23°$ which shows a very narrow line-width of $\Delta\omega = 0.2°$, broad satellite peaks appear symmetrically with respect to the central peak.

Similar rocking curves have been published very recently for manganite films grown on (001) $SrTiO_3$ by Gebhardt et al.[36] and have been discussed in terms of periodic twinning modulation waves and twin domains. For thin films (t = 26 nm) the authors observed satellites with a constant in-plane momentum transfer implying a periodic height modulation with an in-plane periodicity of $D \approx 13$ nm. This has been also found by Jin et al. for $La_{0.7}Sr_{0.3}MnO_3$ on (001) LSAT.[37] For thick films (t = 88 nm), the ω-position of the satellites was found to be independent of the diffraction order indicating individual twin domains with $(001)_{pc}$ lattice planes tilted by an angle $\alpha \approx 0.5°$ with respect to the surface.

In Fig. 5 we show the rocking curves of a thin ($t = 8$ nm) and a thick ($t = 100$ nm) LCO film for different diffraction orders. For the thin film, only the first-order satellite peaks can be observed symmetrically at both sides of the $(00l)_{pc}$ reflection indicating a rather short correlation length. The low intensity of the satellites and the absence of higher-order satellites may be also due to the small sample thickness. The angular difference between the first-order satellites and the $(001)_{pc}$ central peak is $l$-times longer than that between the first-order satellites and the $(00l)_{pc}$ central peaks, evidencing a periodic in-plane modulation. The modulation length is estimated by $D = 1/(2\ \Delta q_x)$ given by Gebhardt,[38] where $\Delta q_x$ is the satellite spacing in reciprocal lattice units. Since we could only observe first-order satellites, we approximated $\Delta q_x$ by the spacing between the central peak and the first-order satellite peak. Thus, the $D$ values given below represent only an estimate of the modulation length. In Fig. 6 a) we plot $D$ as a function of the film thickness. For $t = 8$ nm, $D$ amounts to about 8 nm and increases with increasing $t$ to about 22 nm above $t = 50$ nm, where it seems to saturate. The observed modulation length agrees well with that given in Ref. 36. Hence, the modulation observed by XRD is most likely caused by the periodic lattice modulation of the LCO film. Discrepancies with respect to the TEM data (open symbols in Fig. 6 a)) can be explained by the small regions which are analyzed by TEM imaging.

The rocking curves of the thick film ($t = 100$ nm) show in addition very pronounced satellite peaks at $\alpha \approx \pm 0.65°$, independent of the diffraction order. The peaks can be attributed to macroscopic twin domains which are caused by tilted $(001)_{pc}$ lattice planes with a twinning angle of $\pm 0.65°$. Besides the twin peaks satellite peaks arising from the periodic modulation are present which hints at some relationship between both structures. Fig. 6 b) documents the twinning angle as a function of $t$. Only films with $t > 15$ nm display macroscopic twin domains with $\alpha \approx 0.65°$. The observed twinning angle agrees nearly perfectly with the twinning angle of bulk LCO, $\beta = 90° - \alpha_{pc}$, with the pseudocubic angle $\alpha_{pc} = 90.7°$ and thus documents a relaxation towards the rhombohedral structure.

The *c*-axis lattice parameter determined from the (00*l*) Bragg reflections of $\theta/2\theta$ scans does not change for films in the thickness range 15 nm $\leq t \leq$ 100 nm, i. e., $c_f$ = 3.78 Å. For the in-plane lattice parameter $a_f$ we observe only minor changes with respect to the substrate value $a_s$ = 3.865 Å. Therefore, the pseudo-tetragonal unit-cell volume V = $c_f \times a_f^2 \approx$ 56.5 Å$^3$ is nearly constant for $t \leq$ 100 nm, see Fig. 6 c). A noticeable increase of $c_f$ to 3.80 Å is only observed for films with $t >$ 120 nm. An investigation of the corresponding in-plane film lattice parameters reveals a strong relaxation of $a_f$ as well. The films always show many macroscopic cracks which are possibly caused by the volume shrinkage, twinning and corresponding local strain concentration. The cracking of the films leads to some kind of peeling of the film from the substrate that strongly affects the mosaic spread of the film. Measurements on films with t > 120 nm were therefore discarded for following experiments.

An integration of the rocking curves, shown in Fig. 4, allows us to estimate the volume fraction of the nearly homogeneously strained, pseudomorphic part of the film which is proportional to the area of the central peak at $\omega_0$, and that of the twinned and thus rhombohedrally distorted part of the film which is proportional to the area of the satellite peaks. For the central peak at $\omega_0$ the [001]$_{pc}$ film direction is exactly parallel to the [001] direction of the substrate, so that the narrow peak can be attributed to the pseudomorphic part of the film. The satellite peaks are due to a periodic twinning modulation and nano-twinning and are thus caused by a partial relaxation of strain towards the rhombohedral structure of bulk LCO. Fig. 7 visualizes the decrease of the pseudomorphic (PM) part of the film as a function of *t*. The volume fraction of the PM part decreases almost linearly up to a film thickness $t$ = 15 nm (already at the very beginning, i. e., for films with $t$ = 8 nm). For $t \geq$ 15 nm there is a clear levelling off of the decrease. The functional behavior indicates that the film does not grow completely pseudo-tetragonal up to a certain critical thickness, $t_{cr}$, at which the pseudomorphic growth stops and above which the film starts to relax and to relieve epitaxial strain. For comparison, Fig. 7 also displays the functional behavior for a simple two stage growth model, where in the first stage, for $t \leq t_{cr}$, a pure pseudomorphic and for $t > t_{cr}$ a fully relaxed growth with $t_{cr} = 15$ nm has been assumed. It seems that nano-twinning occurs from the initial stage of growth, however, in an only moderate way by a periodic twinning modulation, whereas pseudomorphic grains still grow well up to at least several tenths of nanometers.

B. Magnetic properties – effective paramagnetic and saturated ferromagnetic moment.

The magnetic properties of the LCO films were characterized by SQUID magnetometry. Fig. 8 displays field-cooled measurements of the magnetic moment, $m$, as a function of temperature for LCO films with different film thickness, $t = 8, 15, 25, 50$ and $100$ nm. Obviously, $m$ increases with increasing $t$ because of the increased amount of magnetic film material. The Curie temperature of the FM phase transition, $T_c$, was determined by extrapolating to $m = 0$. There is a clear reduction of $T_c$ with decreasing $t$. To clarify that, Fig. 9 displays $m/m_0$ versus $T$ for films with different film thickness. $m_0$ corresponds to the magnetic moment at $T = 5$K. Films with $t = 50$ nm show a transition temperature of $T_c \approx 87$ K, whereas for the thinnest film, $t = 8$ nm, we obtain $T_c \approx 80$ K.

The most important factor for a decrease of $T_c$ of thin films with decreasing film thickness is the finite-size effect.[39,40] A scaling analysis of $T_c$ for the LCO films indeed results in a power-law scaling $[T_c(\infty)-T_c(t)]/T_c(\infty) \propto t^{\lambda}$, with a critical shift exponent of $\lambda \approx 1$. Previous investigations on hole-doped cobaltates $La_{0.7}A_{0.3}CoO_3$ (A= Ca, Sr, Ba) as well as on $La_{0.7}Ca_{0.3}MnO_3$ revealed a power-law scaling with the same exponent[41,42,43] which seems to be typical for FM cobaltate and manganite thin films. The rather broad FM transition might likewise be due to a finite-size effect.

Besides the finite-size effect epitaxial strain, structural phase transitions and microstructural deficiencies can cause a reduced $T_c$. However, since the crystalline quality and epitaxial strain can be assumed to increase with decreasing $t$, these effects might be excluded. In addition, previous results have shown that an increase of tensile strain leads to an increase of $T_c$.[4]

The effective paramagnetic moment per Co ion, $m_{eff}$, deduced from the magnetic susceptibility above 100 K is $m_{eff} \approx 3.8$ $\mu_B$/Co. $m_{eff}$ was found to be the same for all LCO films on LSAT in that thickness range and did not depend on $t$. In Fig. 10 we plot the slope $m^* = \partial m/\partial(1/T)$ for $T > 100$ K versus $t$. Assuming a Curie-Weiss law, $m^* \approx nHm_{eff}^2/3k_B$, where $n$ is the number of Co atoms and $k_B$ the Boltzmann constant. Since $n$ is proportional to $t$, the slope of $m^*$ versus $t$ is proportional to $m_{eff}^2$. The linear increase of $m^*$ vs. $t$ demonstrates the constant behavior of $m_{eff}$ over the measured thickness range. The effective paramagnetic moment is determined by the angular moment $J$, i. e., $m_{eff} \propto \sqrt{J(J+1)}$, and therefore strongly related to the spin state of the $Co^{3+}$. Thus, the spin state above 100 K seems to be constant over the measured thickness range.

In contrast, a cracked film with $t = 216$ nm, which contains a significant amount of fully relaxed (bulk like) material, shows a clear reduction of $m^*$ by about 30 % with respect to the expected value from the linear increase of $m^*$ vs. $t$. For bulk LCO $m_{eff}$ amounts to about 2 $\mu_B$/Co[44,45]. Therefore, $m_{eff}$ seems to depend more strongly on the unit-cell volume, which is nearly constant for $t \leq 120$ nm, than on the rhombohedral distortion, which clearly proceeds with increasing $t$. The experimental finding is consistent with generalized-gradient-corrected density functional calculations (GGA+U) by Knížek et al.[12] who determined the dependence of the magnetic state of LCO on crystal structure parameters. Structures without a rhombohedral distortion ($\alpha_{pc} = \eta = 90°$) were compared to analogous structures having the maximum observed rhombohedral deformation of $\eta = 91.5°$, i. e. bulk LCO at $T = 4$ K. The calculations show that the effect of the rhombohedral distortion on the spin state is rather small. The authors found a rather abrupt transition from the LS to the IS state above $T \approx 100$ K where $\eta = 91.4°$.

The O-Co-O bond angle only slightly decreases to η = 91° at 700 K, indicating that the IS state is also stable in the presence of small rhombohedral distortions. A stabilization of the IS state of $Co^{3+}$ was also found for an increasing unit-cell volume by Ravindran *et al.*.[18]

The increased unit-cell volume of the LCO films in comparison to that of bulk LCO, which is nearly constant for 8 nm ≤ $t$ ≤ 100 nm, therefore suggested the reason why $m_{eff}$ does not change for $t$ < 100 nm, even though a rhombohedral distortion is present in some part of the film.

In contrast to the constant $m_{eff}$ in the paramagnetic state above $T_c$, the magnetization in the FM state at 5K and $\mu_0 H$ = 20 mT decreases from about 0.6 to 0.3 $\mu_B$/Co with increasing $t$. The inset of Fig. 8 shows the hysteresis loops of the LCO films, i. e., the magnetic moment $m$ versus the external field strength $\mu_0 H$, at $T$ = 10 K. The coercive field $\mu_0 H_c$ ≈ 0.3 T does not seem to depend significantly on $t$. The saturated moment, $m_s$, the remnant magnetization, $m_r$, and the ratio $m_r/m_s$ all increase with increasing $t$. The plot of $m_s$ versus $t$, see Fig. 10, demonstrates that the increase is sub-linear above $t$ ≈ 25 nm, indicating that some parts of the film exhibit a reduced saturated moment or are even not in the FM state. In the inset of Fig. 10 we plot $m_s$ as a function of $t_{PM}$, where $t_{PM} = t \times$ PM corresponds to an "*effective pseudomorphic thickness*". PM is the volume fraction of the pseudomorphic part of the LCO film, see Fig. 7. Obviously, $m_s$ shows a perfect linear dependence on $t_{PM}$ as well as m* on $t$. Fig. 10 thus demonstrates that the FM state in epitaxially strained LCO films is strongly confined to the pseudo-cubic (tetragonal) phase and instantly suppressed in the presence of a rhombohedral distortion of the unit-cell.

However, the crystal structure of the pseudo-cubic (tetragonal) phase of LCO is less clear. The room-temperature lattice constant of LSAT, $a_s$ = 3.865 Å, is almost exactly twice that of the Co-O bond length measured in bulk LCO at 300 K, $d$ = 1.9345 Å.[24] Assuming that the Co-O bond is rather "hard" this would suggest a Co-O-Co bond angle of almost exactly 180°, i.e., an almost complete "untilting" of the CoO$_6$ octahedra.[4] This would imply in turn that the Co-O-Co bond angle for bonds in $c$-direction must be nearly 180° as well so that the reduction of the $c$-axis lattice constant by about 2% to $c_f$ = 3.78 Å, must be the result of a contraction of the CoO$_6$ octahedra along the $c$-direction. This would still be consistent with EXAFS results.[46] This scenario appears plausible for another reason as well. The rhombohedral distortion in bulk LCO is believed to be caused by the small size of the La$^{3+}$ ion. The rhombohedral distortion reduces the La-O bond length from 2.7358 Å - calculated for the ideal cubic structure with a Co-O bond length of 1.9345 Å - to an average of 2.7148 Å.[24,47] The "ideal" tetragonal structure with all bond angles equal to 180° and Co-O bond lengths of 3.865/2 Å in the $ab$-plane and 3.78/2 Å along the $c$-axis direction indeed gives a La-O distance of 2.7032 Å, i.e,, close to the average distance in the rhombohedral structure.

The GGA+U calculations of Knížek et al. for the rhombohedral structure show that irrespective of the Co-O bond length the magnetic ground state is favored as the Co-O-Co bond angle approaches 180°. This remains true also in the tetragonal phase as shown by GGA calculations and explains the ferromagnetic nature of the films. The electronic and magnetic properties of rhombohedral and tetragonal LCO were investigated by means of electronic structure calculations using density functional theory in the generalized gradient approximation as well as the new full-potential augmented spherical wave method.[48] The energy lowering, $\Delta E$, of the FM state with respect to the nonmagnetic state and the corresponding magnetic moment per formula unit, $m_{uc}$, are given in Tab. I for various hypothetical tetragonal structures (assuming Co-O-Co bond angles of 180°) with lattice parameters $a$ and $c$ that were extracted from LCO films epitaxially grown on different substrate materials.[4]

TABLE I: Total energy lowering $\Delta E$ in mRyd per formula unit and magnetic moment $m_{uc}$ in $\mu_B$ for various lattice parameters $a$ and $c$ in Å and the corresponding tetragonal distortion $TD = a/c$ and cell volume $V$ in Å$^3$.

| a | c | TD | V | ΔE | $m_{uc}$ |
|---|---|---|---|---|---|
| 3.819 | 3.819 | 1.0 0 | 55.306 | -6.88 7 | 1.4 4 |
| 3.803 | 3.869 | 0.98 | 55.956 | -7.56 8 | 1.4 7 |
| 3.855 | 3.811 | 1.0 1 | 56.635 | -8.10 0 | 1.5 0 |
| 3.869 | 3.799 | 1.0 2 | 56.86 7 | -8.55 0 | 1.5 1 |
| 3.889 | 3.790 | 1.03 | 57.321 | -9.291 | 1.53 |

The FM is always lower than the nonmagnetic state by few mRyd where the FM stability increases with increasing cell volume, as does $m_{uc}$. Interestingly, the magnetic moment per volume is very similar to that which has been obtained for the rhombohedral structure and therefore seems to depend more strongly on the cell volume than on the crystal structure. In contrast, $\Delta E$, i. e. the tendency to long-range order does not depend on the volume alone but is significantly influenced by the symmetry. . These calculations confirm the experimental results and are consistent with GGA calculations by Ravindran *et al.* and Knížek *et al.*.

**IV. Conclusions**

Epitaxially strained LCO films grown on (001) LSAT substrates show a nearly pseudomorphic growth, in the initial state of film deposition. This results in a pseudo-tetragonal structure with an increased unit-cell volume in comparison to that of bulk LCO. Up to a film thickness $t \leq 100$ nm, strain is relieved predominantly elastically by nano-twinning along the $<100>_{pc}$ directions, which already starts moderately at the initial stage of growth by the formation of a periodic twinning modulation. On the other hand, pseudomorphic grains still grow up to a thickness of at least several tenths of nanometers. The volume fraction of the pseudomorphic part of the film decreases with increasing $t$ to about 35% for films with $t = 100$ nm. The twinning of the LCO film is attributed to the symmetry lowering of the epitaxially strained pseudo-tetragonal structure towards the relaxed rhombohedral structure of bulk LCO. The unit-cell volume was found to be nearly constant for $t \leq 100$ nm. Films with $t > 120$ nm showed the formation of macroscopic cracks and a significant relaxation of the lattice parameters to values comparable to those of the pseudo-cubic structure of bulk LCO.

The Curie temperature of the LCO films is slightly reduced with decreasing $t$ from $T_c = 87$ K for $t = 50$ nm to $T_c = 80$ K for $t = 8$ nm. The reduction is likely caused by a finite-size effect. The effective paramagnetic moment, $m_{eff}$, and thus the spin state of the $Co^{3+}$ ion was found to be nearly constant for $t \leq 100$ nm. However, the saturated ferromagnetic moment, $m_s$, is proportional only to the pseudo-tetragonal part of the film and decreases with increasing rhombohedral distortion. The measurements demonstrate that the ferromagnetic state of LCO is strongly affected by the rhombohedral distortion and to a much lesser degree by the increased unit-cell volume which mainly controls the effective paramagnetic moment and thus the spin state of the $Co^{3+}$ ion. Epitaxial strain is able to stabilize the pseudo-cubic (tetragonal) structure and thus the FM state until strain relaxation via nano-twinning and the emerging rhombohedral distortion takes place. The experimental results are supported by our density-functional calculations and are consistent with those of Ravindran *et al.* and Knížek *et al.*, thus providing a better understanding of the FM state in LCO thin films.

**Acknowledgement**

This work was partly supported by the DFG in the frame of the Research Unit FO *960* "Quantum Phase Transitions" and through SFB 484. It was also supported by the DFG Research Center for Functional Nanostructures (CFN) within the project F1.2 and by a grant from the Ministry of Science, Research and the Arts of Baden-Württemberg (Az: 7713.14-300).

**Figures and Figure Captions:**

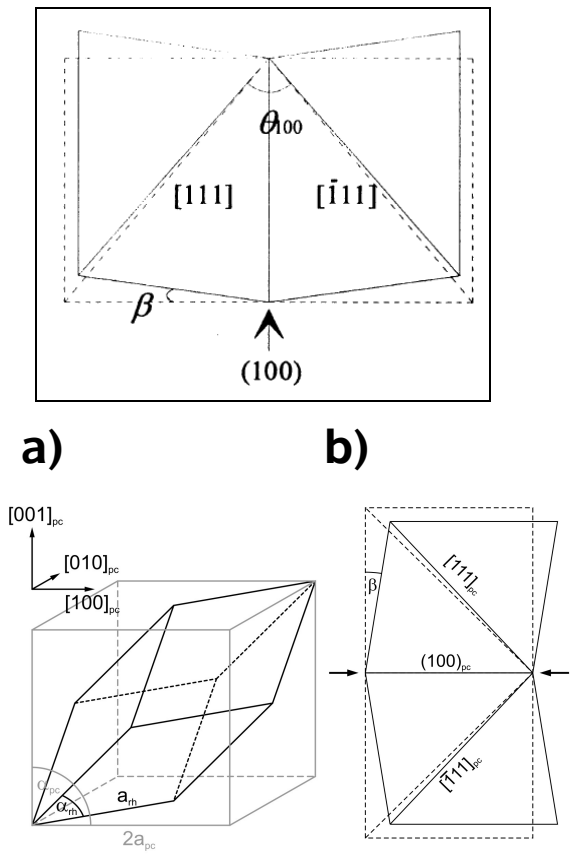

a) b)

FIG. 1. Idealized pseudocubic perovskite structure (grey line) and rhombohedral perovskite structure (black line). The pseudocubic lattice parameter $a_{pc}$, the pseudo-cubic angle $\alpha_{pc}$ and the rhombohedral lattice parameter $a_{rh}$ and angle $\alpha_{rh}$ are shown (a). The geometrical relation between $\alpha_{pc}$ and $\alpha_{rh}$ is given in the text. The rhombohedral phase is formed by a compression along one of the four body diagonals

of the cubic unit cell, <111>$_{pc}$, which results in four different domain states forming (100)$_{pc}$ twin-planes with a twinning angle $\beta = \alpha_{pc} - 90°$ (b).

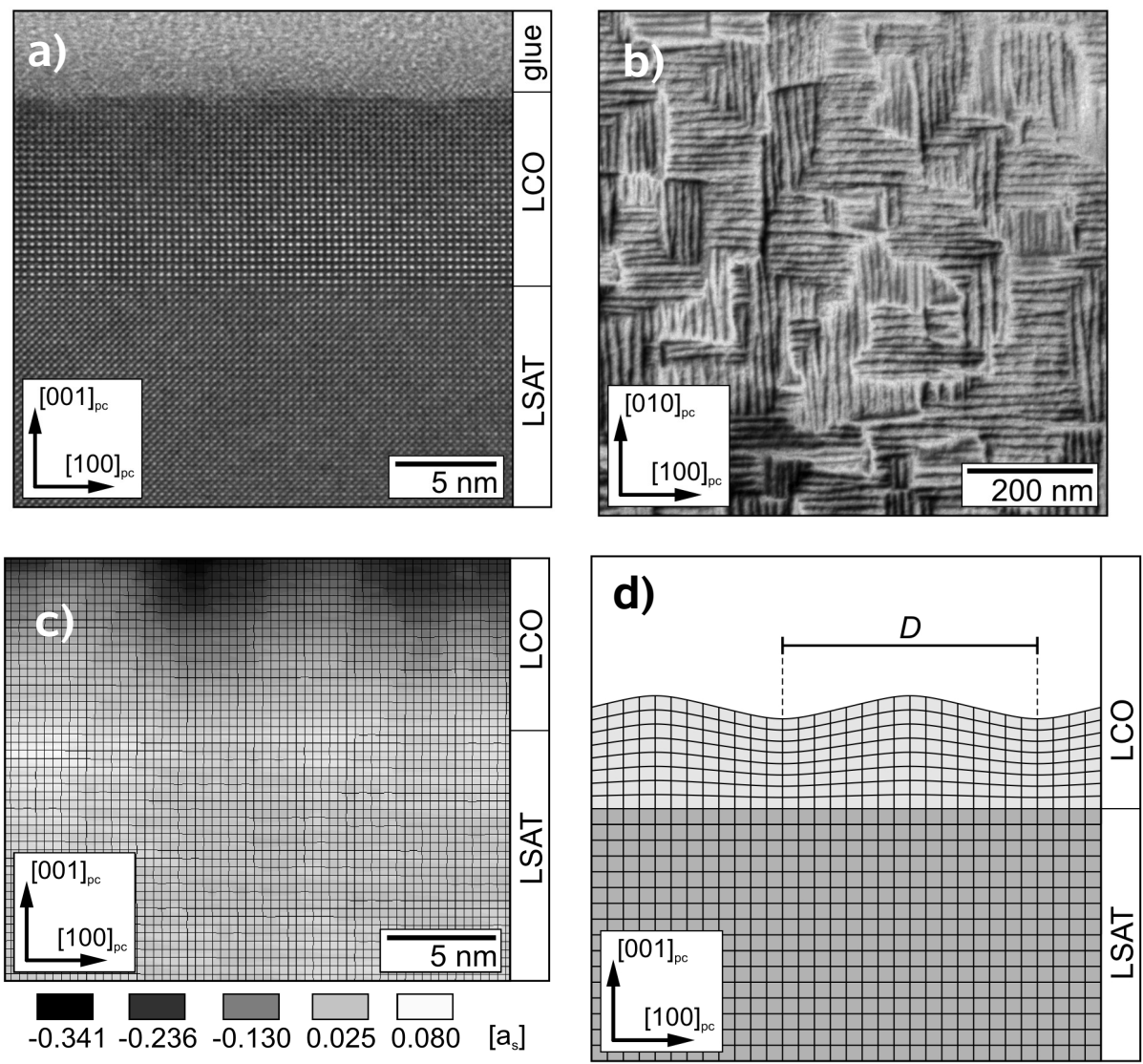

FIG. 2. (a) High-resolution TEM image of an 8 nm LCO film on (001) LSAT imaged along the [010] zone axis; (b) plan-view bright-field image of the 8 nm LCO film imaged under (220)$_{pc}$ two-beam conditions showing domains with periodic lattice modulation along the [100]$_{pc}$ and [010]$_{pc}$ direction;

(c) (color online) Strain-state analysis of Fig. 2(a) showing local displacement of the $(001)_{pc}$ planes with respect to a reference lattice with the lattice parameter of the LSAT substrate, $a_s$. The displacement is given in fractional units of $a_s$. (d) Model of sinusoidally modulated $(001)_{pc}$ lattice planes of LCO with the periodicity $D$.

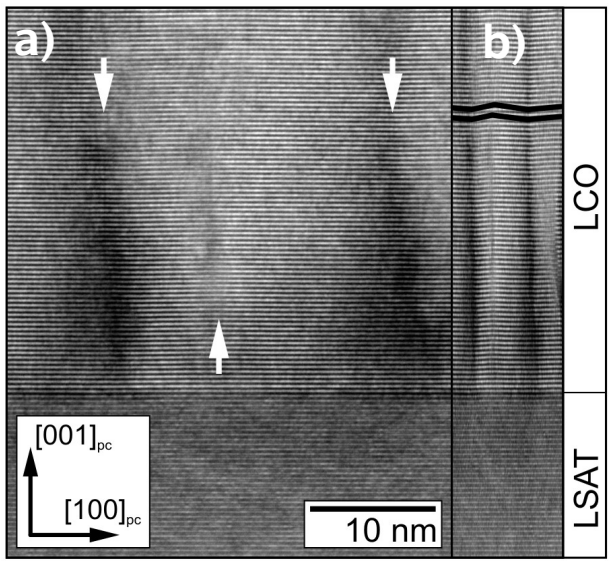

FIG. 3. (a) Cross-section lattice-fringe TEM image of a LCO film ($t \approx 50$ nm) on (001) LSAT viewed along the [010]-zone axis with twin boundaries marked by white arrows; (b) same image as in a) with a scale reduced by a factor of five along the $[100]_{pc}$ direction to emphasize the lattice-plane inclination in the film with respect to the substrate.

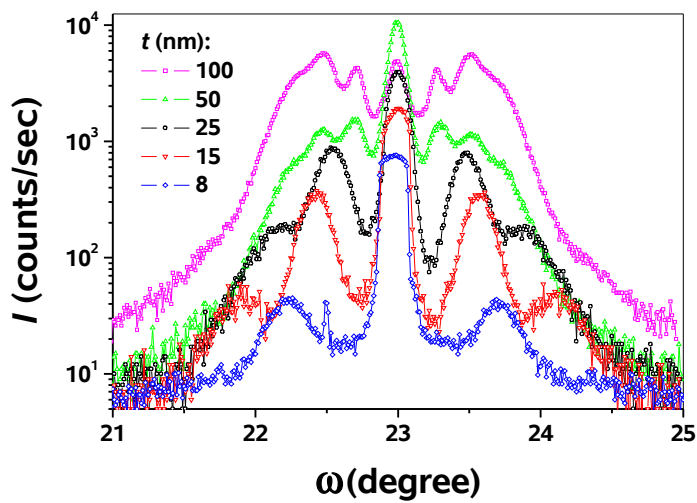

FIG. 4. Rocking curves at the $(002)_{pc}$ reflection of LCO films grown on (001) LSAT for different film thickness, $t$.

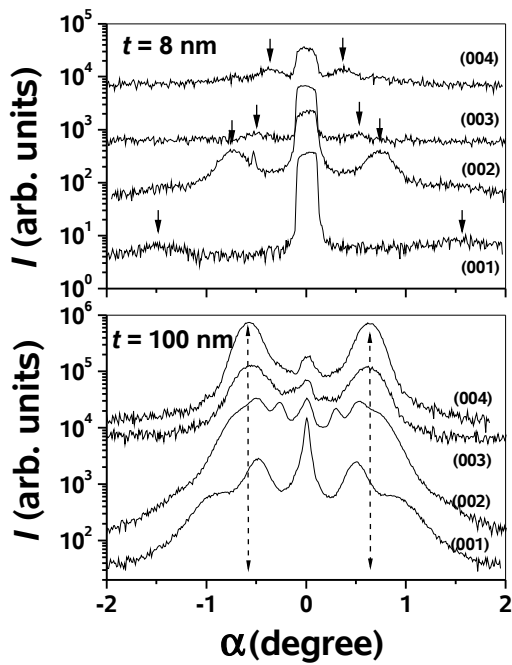

FIG. 5. Rocking curves of a thin ($t$ = 8 nm) and a thick ($t$ = 100 nm) LCO film for different orders of Bragg reflections. The intensity is plotted as a function of $\alpha = \omega - \omega_0$, where $\omega_0$ corresponds to the position of the central peak. Satellite peaks caused by periodic twinning modulation and lattice plane tilt are marked by arrows.

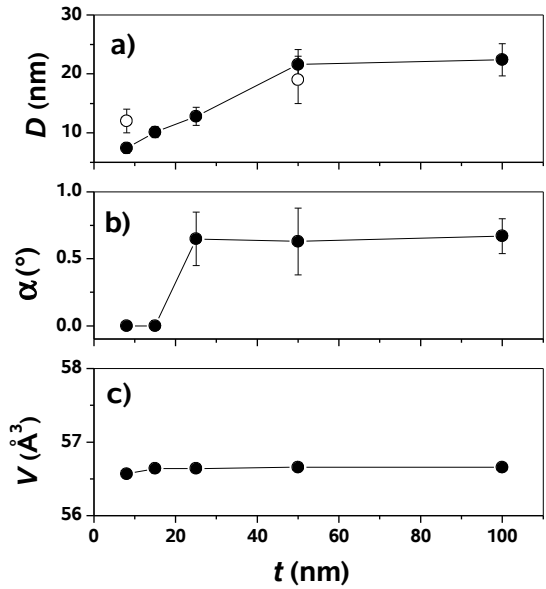

FIG. 6. The modulation length $D$ of the periodic in-plane structure (a) and the domain twinning angle $\alpha$ (b) as determined from rocking curves versus the film thickness $t$. $D$ values as extracted from TEM measurements are shown by open symbols. The pseudo-tetragonal unit cell volume $V = c_f \times a_f^2$ of LCO films on (001) LSAT versus $t$ (c).

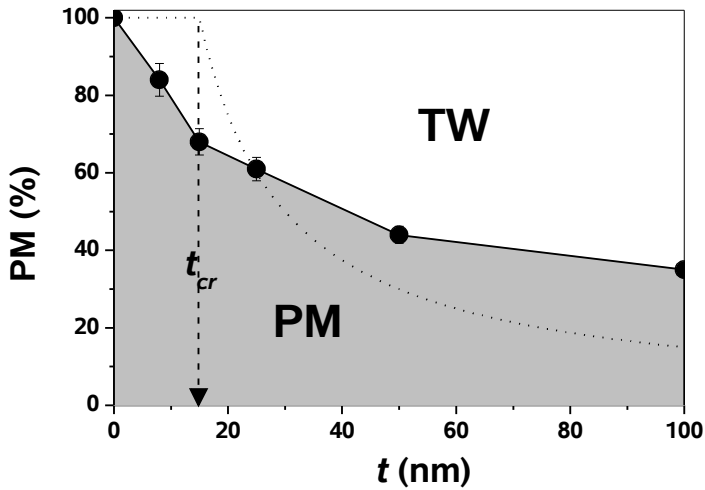

FIG. 7. The volume fraction of the pseudomorphic (PM) part of the LCO film versus the film thickness $t$. The grey and white coloured area correspond to the PM and twinned (TW) part of the film. The dotted line indicates the functional behavior for a simple two stage growth model, where in the first stage, for $t \leq t_{cr}$, a pure pseudomorphic and for $t > t_{cr}$ a fully relaxed growth with $t_{cr} = 15$ nm has been assumed.

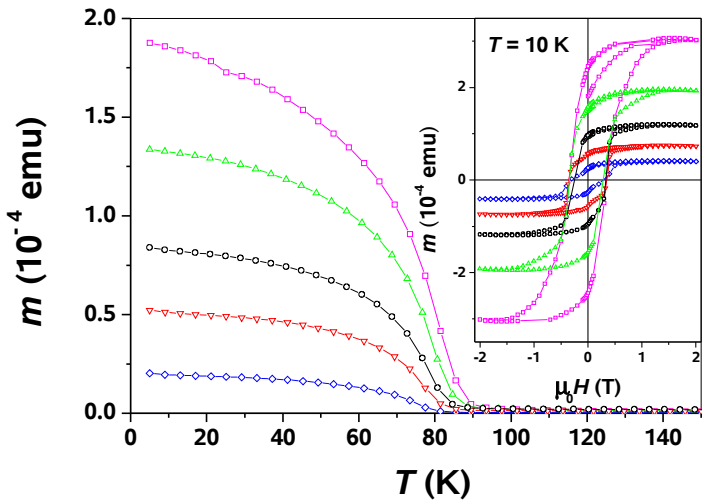

FIG. 8. Magnetic moment, $m$, as a function of $T$ for LCO films with different film thickness, $t = 8, 15, 25, 50,$ and $100$ nm), increasing from bottom to the top, respectively. The field-cooled measurements

were carried out with a field strength of $\mu_0 H = 20$ mT, applied parallel to the film surface. The inset shows the hysteresis loops of the corresponding samples, $m$ versus $\mu_0 H$ at $T = 10$ K. The saturated moment, $m_s$, is increasing with increasing $t$.

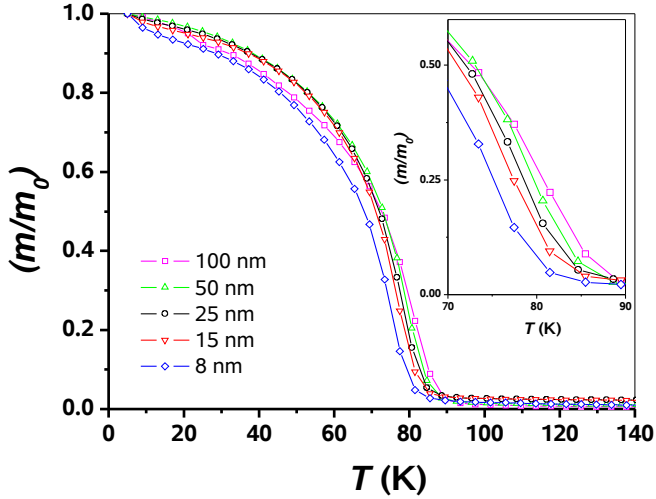

FIG. 9. $m/m_0$ versus $T$ for films with different film thickness. $m_0$ corresponds to the magnetic moment at $T = 5$K.

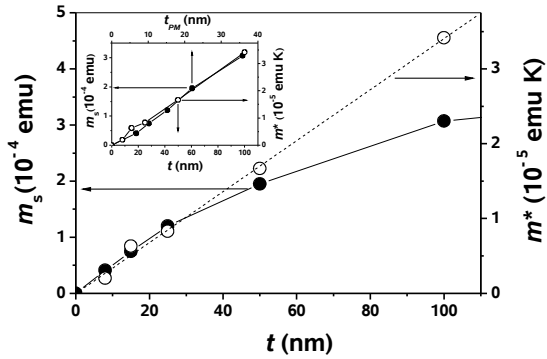

FIG.10. Saturated ferromagnetic moment, $m_s$, at $T = 10$ K (closed symbols, left scale) and $m^* = \partial m/\partial(1/T)$ (open symbols, right scale), which is proportional to the square of the effective paramagnetic moment above $T_c$ and the film thickness, i. e., $m^* \propto t \times m_{\text{eff}}^2$, shown as a function of the film thickness $t$. The inset demonstrates the linear behaviour of $m^*$ and $m_s$ versus $t$ and $t_{PM}$,

respectively, where $t_{PM} = t \times PM$ corresponds to an "effective pseudomorphic thickness". PM is the volume fraction of the pseudomorphic part of the LCO film, see Fig. 7.